\documentclass[twocolumn,preprintnumbers,amsmath,amssymb,prb]{revtex4-1}

\usepackage{graphicx}
\usepackage{dcolumn}
\usepackage{subfigure}
\usepackage{color}
\usepackage{bm}

\begin{document}

\title{Structure and lattice dynamics of the wide band gap semiconductors MgSiN$_{2}$ and MgGeN$_{2}$}

\author{M. R{\aa}sander}\email{m.rasander@imperial.ac.uk}
\affiliation{%
Department of Materials, Imperial College London, SW7 2AZ, London, United Kingdom
}%
\author{J. B. Quirk}
\affiliation{%
Department of Materials, Imperial College London, SW7 2AZ, London, United Kingdom
}%
\author{T. Wang}
\affiliation{%
Department of Materials, Imperial College London, SW7 2AZ, London, United Kingdom
}%
\author{S. Mathew}
\affiliation{%
Department of Chemistry, University College London, Gordon Street WC1H 0AJ, United Kingdom
}%
\author{R. Davies}
\affiliation{%
Department of Materials, Imperial College London, SW7 2AZ, London, United Kingdom
}%
\author{R. Palgrave}
\affiliation{%
Department of Chemistry, University College London, Gordon Street WC1H 0AJ, United Kingdom
}%
\author{M. A. Moram}
\affiliation{%
Department of Materials, Imperial College London, SW7 2AZ, London, United Kingdom
}%
\date{\today}

\begin{abstract}
We have determined the structural and lattice dynamical properties of the orthorhombic, wide band gap semiconductors MgSiN$_{2}$ and MgGeN$_{2}$ using density functional theory. In addition, we present structural properties and Raman spectra of a MgSiN$_{2}$ powder. The structural properties and lattice dynamics of the orthorhombic systems are compared to wurtzite AlN. We find clear differences in the lattice dynamics between MgSiN$_{2}$, MgGeN$_{2}$ and AlN, for example we find that the highest phonon frequency in MgSiN$_{2}$ is about 100~cm$^{-1}$ higher than the highest frequency in AlN and that MgGeN$_{2}$ is much softer. We also provide the Born effective charge tensors and dielectric tensors of MgSiN$_{2}$, MgGeN$_{2}$ and AlN. Phonon related thermodynamic properties, such as the heat capacity and entropy, are in very good agreement with available experimental results.
\end{abstract}

\maketitle

\section{Introduction}
The Group III-nitride semiconductors AlN, GaN and InN are used widely in optoelectronics and high power electronic devices.\cite{a,b,c} AlN and its alloys are also used widely in energy harvesting devices and RF applications.\cite{d} However, improved efficiencies are required for III-nitride-based ultraviolet (UV) light emitting diodes\cite{e}, solar cells\cite{f} and energy harvesting devices.\cite{d} Group II-IV nitride semiconductors are of growing interest in this regard, as their bonding and crystal structures are related to those of III-nitrides but they offer different combinations of band gaps and lattice parameters, opening up additional possibilities for device design.\cite{g,h} For example, Zn-based II-IV nitrides are of current interest for solar cells,\cite{g,h} whereas wide band gap II-IV nitrides, such as MgSiN$_{2}$, may find applications as part of UV optoelectronic devices.\cite{Quirk,Rasander2016,Punya2016} 
\par
MgSiN$_{2}$ and MgGeN$_{2}$ have been found to possess a wurtzite-derived orthorhombic crystal structure which belongs to the Pna2$_{1}$ space group (No. 33).\cite{David1970} The band gap of MgSiN$_{2}$ has been investigated by both theory\cite{Quirk,Rasander2016,Punya2016,Fang1,Huang,deBoer2015} and experiment,\cite{Quirk,deBoer2015,Gaido1974} and it has been found that the band gap in MgSiN$_{2}$ is in-direct with a similar size as the band gap in wurtzite AlN.\cite{Quirk,Punya2016,deBoer2015} A recent study employing the GW approximation has established that the band gap decreases from 5.85~eV in MgSiN$_{2}$ to 5.14~eV in MgGeN$_{2}$ to 3.43~eV in MgSnN$_{2}$ when the Group IV element is substituted from Si to Ge to Sn.\cite{Punya2016} It was also found that the band gaps in both MgGeN$_{2}$ and MgSnN$_{2}$ are direct, in contrast to the indirect gap in MgSiN$_{2}$. A similar trend is also observed for Zn-IV-N$_{2}$ nitride phases,\cite{g,h} however, the sizes of the band gaps are smaller in these Zn-based systems than in the equivalent Mg-based II-IV nitrides.\cite{g,h} In addition to having a large band gap, MgSiN$_{2}$ has been found to possess a small anisotropic thermal expansion\cite{Bruls1} and a thermal conductivity of 28~W/mK obtained for a powder at 300~K.\cite{Bruls2005} These properties together with the great lattice matching between MgSiN$_{2}$ and AlN\cite{Quirk} suggest that MgSiN$_{2}$ could be used for short wavelength or high power electronic devices where heat transport away from the active region of the device is highly desirable.
\par
In this study we present structural as well as phonon-related properties, such as phonon dispersions, phonon density of states, the Helmholtz free energy and the heat capacity, of MgSiN$_{2}$ and MgGeN$_{2}$, which has not been done previously, even though there is an experimental study of the heat capacity of MgSiN$_{2}$.\cite{Bruls1998} The bulk of the results presented here are obtained using calculations based on density functional theory. However, we also provide experimental structural properties and Raman spectra of a MgSiN$_{2}$ powder. In order to facilitate greater understanding, the properties established for MgSiN$_{2}$ and MgGeN$_{2}$ will be compared to wurtzite AlN, which is the starting point for obtaining the crystal structure of MgSiN$_{2}$. We note that both MgSiN$_{2}$ and MgGeN$_{2}$ have been produced in powder form previously,\cite{David1970} and, considering the size of the band gaps,\cite{Punya2016} are of interest in UV-optoelectronic applications.

\section{Method}
\par
Density functional calculations have been performed using the projector augmented wave (PAW) method\cite{Blochl} as implemented in the Vienna {\it ab initio} simulations package (VASP).\cite{KresseandFurth,KresseandJoubert} We have used the generalized gradient approximation of Perdew, Burke and Ernzerhof (PBE)\cite{PBE} for the exchange-correlation energy functional. The plane wave energy cut-off was set to 800~eV and we have used $\Gamma$-centered k-point meshes with the smallest allowed spacing between k-points of 0.1~\AA$^{-1}$. The atomic positions and simulation cell shapes were relaxed until the Hellmann-Feynman forces acting on atoms were smaller than 0.0001~eV/\AA. Furthermore, we have used the standard core-valence partitioning for Mg (with a 3s$^{2}$ valence), Si (with a 3s$^2$3p$^2$ valence), Al (with a 3s$^2$3p$^1$ valence) and N (with a 2s$^2$2p$^3$ valence). In the case of Ge we have used both the four 4s and 4p electrons and the semi-core 3d electrons as valence states.
\par
The evaluation of phonon related properties have been obtained using the finite displacement method as it is implemented in the Phonopy code.\cite{Chaput2011,Parlinski1997} For MgSiN$_{2}$ and MgGeN$_{2}$ we have used supercells based on a 3$\times$3$\times$3 repetition of the primitive orthorhombic unit cells containing 432 atoms. In the case of AlN, we have used a 5$\times$5$\times$4 repetition of the primitive wurtzite unit cell with 400 atoms in the supercell. The phonon density of states and related properties were obtained using a 24$\times$24$\times$24 q-point mesh for MgSiN$_{2}$ and MgGeN$_{2}$ and a 48$\times$48$\times$48 q-point mesh for AlN.
\par
In polar crystals, a macroscopic electric field is induced by the collective displacements of the ions at ${\bm q}=0$. This electric field affects the longitudinal optical (LO) modes and not the transverse optical (TO) modes and therefore gives rise to a LO-TO splitting of the modes when approaching the $\Gamma$ point. To account for this effect, a non-analytical correction to the dynamical matrix on the form 
\begin{equation}\label{eq:D-corr}
D_{\alpha j,\beta j'}^{NA}({\bm q}\rightarrow0) = \frac{4\pi}{\Omega_{0}\sqrt{m_{j}m_{j'}}} \frac{\left[ \sum_{\gamma}q_{\gamma}{\bm Z}^{\ast}_{j,\gamma\alpha} \right]\left[ \sum_{\gamma'}q_{\gamma'}{\bm Z}^{\ast}_{j',\gamma'\beta} \right]}{\sum_{\alpha\beta}q_{\alpha}{\bm\varepsilon}_{\alpha\beta}^{\infty}q_{\beta}}
\end{equation}
is added.\cite{Pick1970,Giannozzi1991,Gonze1997,Wang2010} Here ${\bm\varepsilon}^{\infty}$ is the high-frequency dielectric tensor and ${\bm Z}_{s}^{\ast}$ is the Born effective charge tensor for the atom $s$.
\par
In addition, MgSiN$_{2}$ powders were synthesised by firing stoichiometric Mg and Si$_{3}$N$_{4}$ powders in a horizontal tube furnace for 16 hours at 1000$^\circ$C using a heating rate of 10$^\circ$C$\cdot$min$^{-1}$ under a flowing N$_{2}$ (700 sccm) atmosphere in a Mo boat, prior to cooling at 5$^\circ$C$\cdot$min$^{-1}$. To verify the crystal structure, powder X-ray diffraction (XRD) was performed over the range 20$^\circ$-120$^\circ$ $2\theta$ using Cu-K$\alpha_{1}$ radiation. Furthermore, Raman measurements were collected using a LabRAM HR Evolution spectrometer (Horiba Scientific) using laser wavelengths of 532 nm, an objective lens of $\times$10, and a grating of 600 lines/mm.  Spectra were recorded over the range 100 to 1250 cm$^{-1}$ wavenumbers using an acquisition time of 100 s.
\par
\begin{figure}
\includegraphics[width=10cm]{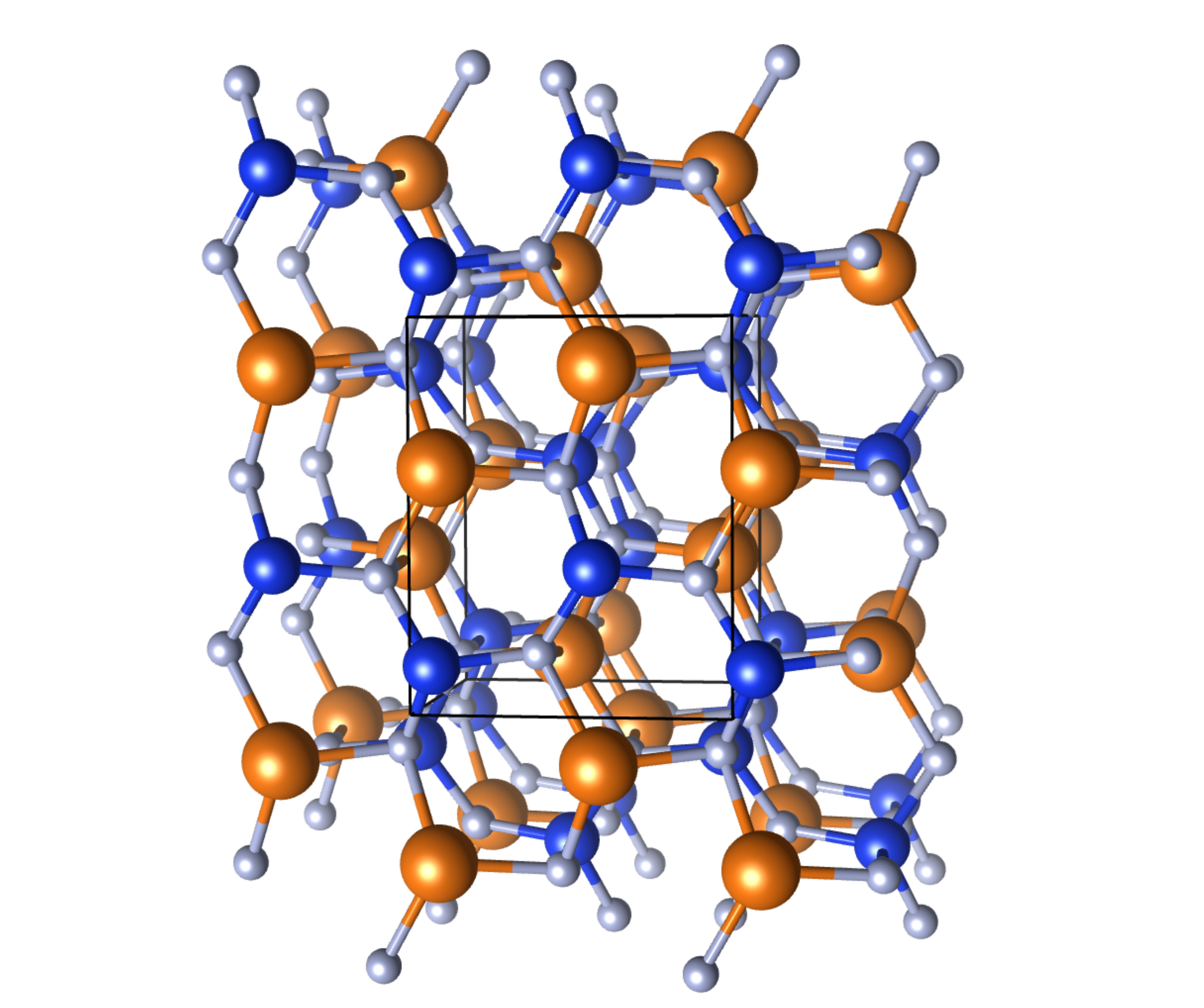}
\caption{\label{fig:planes} (Color online) The orthorhombic crystal structure (space group Pna2$_{1}$) of MgSiN$_{2}$ and MgGeN$_{2}$. Mg is represented by the large bronze colored spheres, Si (or Ge) is represented by blue spheres, while N is represented by the small grey spheres.}
\end{figure}
\begin{table}[b]
\caption{\label{tab:structure-MgSiN2} Lattice constants and crystallographic coordinates $(x,y,z)$ of orthorhombic MgSiN$_{2}$. The numbers in parenthesis are the deviation from the experimental value.}
\begin{ruledtabular}
\begin{tabular}{lccc}
XC & $a$ (\AA) & $b$ (\AA) & $c$ (\AA) \\
 \hline
 PBE & 5.311 (0.8\%) & 6.495 (0.4\%)& 5.028 (0.9\%) \\
 Expt. ($T=300$~K) & 5.314 & 6.466 & 4.975\\
 Expt. ($T=10$~K)\cite{Bruls1} & 5.27078(5) & 6.46916(7) & 4.98401(5) \\
 \hline
   \multicolumn{4}{c}{PBE} \\
  & $x$ & $y$ & $z$ \\
 \hline
 Mg & 0.0849 &  0.6228 &  0.9886\\
 Si & 0.0700 & 0.1254 &  0.0000\\
 N(1) & 0.0485 & 0.0957 & 0.3471\\
 N(2) & 0.1093 & 0.6553 & 0.4102\\
 \hline
 \multicolumn{4}{c}{Expt. ($T=10$~K)\cite{Bruls1}}\\
& $x$ & $y$ & $z$ \\
\hline
Mg & 0.08448(34) & 0.62255(30) & 0.9866(5) \\
Si & 0.0693(5) & 0.1249(4) & 0.0000\\
N(1) & 0.04855(17) & 0.09562(15) & 0.3472(4)\\
N(2) & 0.10859(18) & 0.65527(14) & 0.4102(4)\\
 \end{tabular}
\end{ruledtabular}
\end{table}
\begin{table}[bt]
\caption{\label{tab:structure-MgGeN2} Lattice constants and crystallographic coordinates $(x,y,z)$ of orthorhombic MgGeN$_{2}$. The numbers in parenthesis are the deviation from the experimental value.}
\begin{ruledtabular}
\begin{tabular}{lccc}
XC & $a$ (\AA) & $b$ (\AA) & $c$ (\AA) \\
 \hline
 PBE & 5.549 (1.0\%)& 6.658 (0.7\%) & 5.223 (1.1\%) \\
 Expt.\cite{David1970} & 5.494 & 6.611 & 5.166 \\
 \hline
   \multicolumn{4}{c}{PBE} \\
  & $x$ & $y$ & $z$ \\
 \hline
 Mg & 0.0851 & 0.6234 & 0.9929 \\
Ge & 0.0736 & 0.1258 & 0.0000 \\
 N(1) & 0.0604 & 0.1073 & 0.3600 \\
 N(2) & 0.1002 & 0.6431 & 0.3975\\
 \hline
  \multicolumn{4}{c}{Expt.\cite{David1970}}\\
& $x$ & $y$ & $z$ \\
\hline
Mg & 0.083 & 0.625 & 0.000 \\
Ge & 0.083 & 0.125 & 0.000\\
N(1) & 0.083 & 0.125  & 0.380\\
N(2) & 0.083 & 0.625 & 0.400\\
 \end{tabular}
\end{ruledtabular}
\end{table}

\section{Structural properties}\label{sec:structure}

The ordered MgSiN$_{2}$ and MgGeN$_{2}$ structures can be derived from the AlN structure by substituting one Mg and one Si (or Ge) atom for every two Al atoms. The structures have been found to be orthorhombic and belong to the Pna2$_{1}$ space group (No. 33),\cite{David1970,Bruls1} where all atoms occupy the $4a$ wyckoff crystal positions and the unit cell, therefore, contains 4 formula units of MgSiN$_{2}$ or MgGeN$_{2}$. The crystal structure is shown in Fig.~\ref{fig:planes} and the structural parameters are presented in Table~\ref{tab:structure-MgSiN2} for MgSiN$_{2}$ and in Table~\ref{tab:structure-MgGeN2} for MgGeN$_{2}$. We note that in this structure there are two different nitrogen positions, labelled N(1) and N(2). The N(1) positions is approximately positioned above the Si (or Ge) positions along the $c$-axis, while the N(2) positions is approximately positioned above the Mg positions. In order to satisfy local charge neutrality, each N(1) and N(2) position has two Mg and two Si (or Ge) atoms as nearest neighbours. Correspondingly, each Mg and Si (or Ge) position has two N(1) and two N(2) positions as nearest neighbours.
\begin{figure}[b]
\includegraphics[width=8cm]{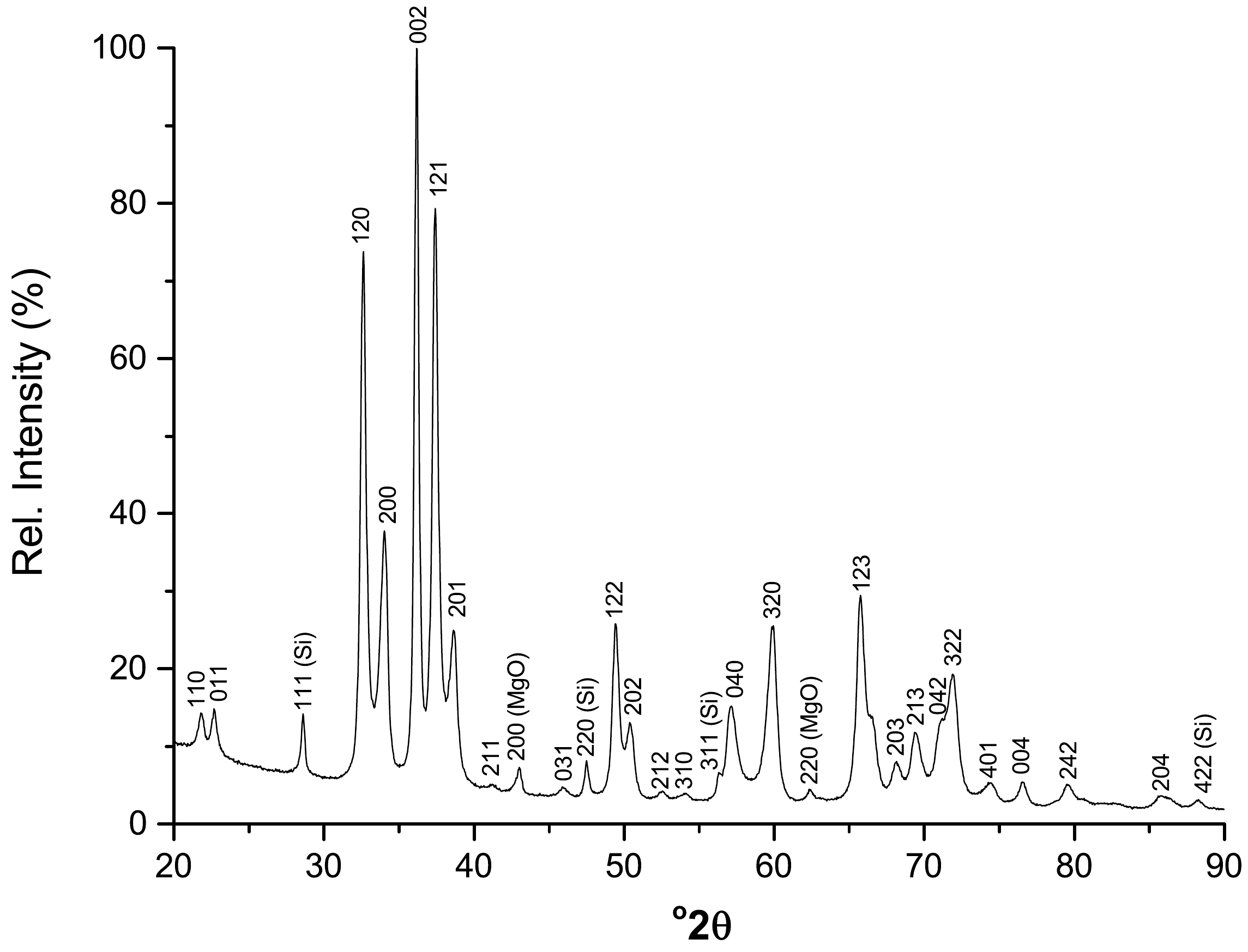}
\caption{\label{fig:xrd} Measured X-ray diffractogram of MgSiN$_{2}$ powder, indexed as the ordered orthorhombic structure belonging to the Pna2$_{1}$ space group.}
\end{figure}
\par
In Fig.~\ref{fig:xrd} we show the measured XRD pattern of MgSiN$_{2}$, which shows that the structure is orthorhombic and belongs to the Pna2$_{1}$ space group. We also find peaks related to pure Si and MgO impurities in the sample. The corresponding MgSiN$_{2}$ lattice constants are shown in Table~\ref{tab:structure-MgSiN2}. The measured lattice constants are in very good agreement with previously measured lattice constants. 
\par
As shown in Table~\ref{tab:structure-MgSiN2} and \ref{tab:structure-MgGeN2}, we find that the PBE calculations overestimate the lattice constants compared to available experiments, with a maximum deviation of 0.9\% for MgSiN$_{2}$ and 1.1\% for MgGeN$_{2}$, which is to be expected using the PBE approximation for these types of systems.\cite{Rasander2015} The calculated structural parameters are also in agreement with previous results obtained using local and semi-local approximations for the exchange-correlation energy functional.\cite{Punya2016,Fang1,Huang}  We note that MgGeN$_{2}$ is a larger crystal than MgSiN$_{2}$, but due to the larger mass of Ge compared to Si the mass density of MgGeN$_{2}$ is larger than in MgSiN$_{2}$. In the case of wurtzite AlN, the calculated structural parameters are $a=3.127$~\AA, $c=5.013$~\AA~and $u=0.382$, which are in good agreement with experiment. When comparing the volume of MgSiN$_{2}$ with the same number of AlN atoms, we find that the volume of MgSiN$_{2}$ is slightly larger at 173.44~\AA$^3$ compared to 169.80~\AA$^3$ in the case of AlN which leads to the mass density being larger in AlN than in MgSiN$_{2}$.
\par
In MgSiN$_{2}$ and MgGeN$_{2}$, all atoms are locally found in a distorted tetrahedral environment, which has been discussed in detail by Bruls {\it et al.}\cite{Bruls1} For MgSiN$_{2}$, we obtain average Mg-N and Si-N binding distances of 2.11~\AA~and 1.76~\AA, respectively, which is slightly larger than the 2.09~\AA~and 1.75~\AA~obtained by Bruls {\it et al.}\cite{Bruls1} In the case of MgGeN$_{2}$, we obtain Mg-N and Ge-N binding distances of 2.10~\AA~and 1.89~\AA, respectively, i.e., the average Mg-N binding distance in MgGeN$_{2}$ is only slightly longer than in MgSiN$_{2}$ whereas the average IV-N binding distance is significantly longer in MgGeN$_{2}$.  In AlN, the average Al-N binding distance is 1.90~\AA. If we average over the Mg-N and Si-N binding distances in MgSiN$_{2}$ the resulting average binding distance is 1.94~\AA, which is slightly larger than the binding distance in AlN, and if we average over the Mg-N and Ge-N binding distances we find the average binding distance in MgGeN$_{2}$ to be 2.00~\AA.
\par
The orthorhombic structure discussed here is derived from the wurtzite crystal structure by transforming the lattice vectors in the $xy$-plane as ${\bm a}={\bm a}_{1}+2{\bm a}_{2}$ and ${\bm b}=2{\bm a}_{1}$ while keeping the ${\bm c}$ lattice vector the same. Here ${\bm a}_{1}$ and ${\bm a}_{2}$ are the in-plane lattice vectors of the wurtzite crystal structure while ${\bm a}$ and ${\bm b}$ are the in-plane lattice constants in the orthorhombic structure. The relation between the in-plane lattice constant in wurtzite, $a_{w}$, and the $a$ and $b$ lattice constants of the orthorhombic structure is therefore $a_{w}\approx a/\sqrt{3}\approx b/2$. It is therefore possible to evaluate an average wurtzite-like lattice constant for the orthorhombic structures as $\bar{a}_{w} = (a/\sqrt{3}+b/2)/2$. For MgSiN$_{2}$ the wurtzite-like lattice constant is 3.157~\AA~which is only slightly larger than the in-plane lattice constant of wurtzite AlN of 3.127~\AA. The corresponding average lattice constant in MgGeN$_{2}$ is 3.267~\AA~which is slightly larger than what is obtained within the PBE approximation for wurtzite GaN (3.218~\AA). Regarding the out-of-plane lattice constants along the $z$-direction, we find the $c$ lattice constant in MgSiN$_{2}$ to be slightly larger that the $c$ lattice constant in AlN, 5.028~\AA~versus 5.013~\AA, respectively. In MgGeN$_{2}$ the $c$ lattice constant is larger and more similar in size to the corresponding lattice constant in GaN, 5.223~\AA~in MgGeN$_{2}$ versus 5.243~\AA~in GaN. It is therefore clear that the relative volume and other structural properties increase going from AlN to MgSiN$_{2}$ to MgGeN$_{2}$, and, furthermore, while the difference between AlN and MgSiN$_{2}$ is rather small, MgGeN$_{2}$ is structurally more similar to wurtzite GaN. 
\begin{table}[t]
\caption{\label{tab:dielectric_tensor} The dielectric tensor, ${\bm \varepsilon}_{\infty}$, MgSiN$_{2}$ and wurtzite AlN. Note that the ${\bm \varepsilon}_{xx}$ component for AlN is the in-plane dielectric constant, while the ${\bm \varepsilon}_{zz}$ component is the out-of-plane dielectric constant along the $c$-axis.}
\begin{ruledtabular}
\begin{tabular}{lccc}
 & ${\bm \varepsilon}_{xx}$ & ${\bm \varepsilon}_{yy}$ & ${\bm \varepsilon}_{zz}$  \\
 \hline
 MgSiN$_{2}$ & 4.37 & 4.29 & 4.42 \\
 MgGeN$_{2}$ & 5.12 & 4.88 & 5.09\\
 AlN & 4.47 & - & 4.70 \\
 AlN\cite{Bungaro2000} & 5.17 & -& 5.36 \\
 AlN\cite{Karch1997} & 4.38 & - & 4.61\\
 \end{tabular}
\end{ruledtabular}
\end{table}


\begin{table}[b]
\caption{\label{tab:Borncharges-MgSiN2} The Born effective charge tensor ${\bm Z}^{\ast}$ for MgSiN$_{2}$.}
\begin{ruledtabular}
\begin{tabular}{lcccc}
 & Mg & Si & N(1) & N(2) \\
 \hline
${\bm Z}_{xx}^{\ast}$ & 1.92 & 3.17 & -2.04 & -3.04\\
${\bm Z}_{xy}^{\ast}$ & 0.04 & -0.19 & -0.11 & 0.47\\
${\bm Z}_{yx}^{\ast}$ & -0.04 & 0.19 &-0.30 & 0.51\\ 
${\bm Z}_{yy}^{\ast}$ & 1.95 & 3.02 & -2.50 & -2.47\\
${\bm Z}_{yz}^{\ast}$ & 0.01 & -0.20 & 0.49 & -0.15\\
${\bm Z}_{zy}^{\ast}$ & -0.04 & 0.23 & 0.49 & -0.13\\
 ${\bm Z}_{zz}^{\ast}$ & 2.08 & 3.12 & -3.04 & -2.16\\
 ${\bm Z}_{xz}^{\ast}$ & -0.02 & -0.05 & 0.17 & -0.10\\ 
 ${\bm Z}_{zx}^{\ast}$ & -0.03 & 0.09 & 0.19 & -0.04\\ 
 \end{tabular}
\end{ruledtabular}
\end{table}

\begin{table}[t]
\caption{\label{tab:Borncharges-MgGeN2} The Born effective charge tensor ${\bm Z}^{\ast}$ for MgGeN$_{2}$.}
\begin{ruledtabular}
\begin{tabular}{lcccc}
 & Mg & Ge & N(1) & N(2) \\
 \hline
${\bm Z}_{xx}^{\ast}$ & 1.86 & 3.18 & -2.07 & -2.98 \\
${\bm Z}_{xy}^{\ast}$ & 0.16 & -0.44 & -0.04 & 0.34\\
${\bm Z}_{yx}^{\ast}$ & -0.07 & 0.31 & -0.33 & 0.47\\ 
${\bm Z}_{yy}^{\ast}$ & 1.97 & 2.96 & -2.48 & -2.45\\
${\bm Z}_{yz}^{\ast}$ & 0.08 & -0.44 & 0.41 & -0.11\\
${\bm Z}_{zy}^{\ast}$ & -0.11 & 0.38 & 0.42 & -0.10\\
 ${\bm Z}_{zz}^{\ast}$ & 2.02 & 3.15 & -2.97 & -2.21\\
 ${\bm Z}_{xz}^{\ast}$ & 0.01 & -0.09 & 0.14 & -0.10\\ 
 ${\bm Z}_{zx}^{\ast}$ & -0.08 & 0.19 & 0.16 & -0.02 \\ 
 \end{tabular}
\end{ruledtabular}
\end{table}

\begin{table}[t]
\caption{\label{tab:Borncharges-AlN} The Born effective charge tensor ${\bm Z}^{\ast}$ for wurtzite AlN.}
\begin{ruledtabular}
\begin{tabular}{lccc}
 & ${\bm Z}_{xx}^{\ast}$ & ${\bm Z}_{yy}^{\ast}$ & ${\bm Z}_{zz}^{\ast}$\\
 \hline
 Al (N) & 2.53 & 2.53 & 2.69\\
Al (N)\cite{Bungaro2000} & 2.70 & 2.70 & 2.85 \\
Al (N)\cite{Karch1997}  & 2.63 & 2.53 & 2.69 \\
 \end{tabular}
\end{ruledtabular}
\end{table}

\section{Dielectric tensors and Born effective charges}
\par
The dielectric tensors of MgSiN$_{2}$, MgGeN$_{2}$ and AlN are shown in Table~\ref{tab:dielectric_tensor}. In the case of MgSiN$_{2}$ and MgGeN$_{2}$, we find the dielectric tensor to be anisotropic, with the largest component along the $c$-axis and the smallest component along the $b$-axis in the case of MgSiN$_{2}$. For MgGeN$_{2}$ the largest component is along the $a$-axis while the smallest component is along the $b$-axis. In both these cases the $a$- and $c$-components are similar in size while the $b$-component is much smaller. The absolute values of the dielectric tensor are larger in MgGeN$_{2}$ compared to MgSiN$_{2}$ which is related to the reduced band gap in MgGeN$_{2}$. When comparing the dielectric tensor of MgSiN$_{2}$ to wurtzite AlN we find both the $a$- and $b$-components to be smaller than the in-plane dielectric constant in AlN. Furthermore, we also find the $c$-component in MgSiN$_{2}$ to be smaller then the out-of-plane component in AlN. In Table~\ref{tab:dielectric_tensor}, we also show the dielectric tensors of AlN obtained in earlier computational studies and our calculations are in good agreement with these previous studies. 
\par
The Born effective charge tensors in MgSiN$_{2}$, MgGeN$_{2}$ and AlN are shown in Tables~\ref{tab:Borncharges-MgSiN2}, \ref{tab:Borncharges-MgGeN2} and \ref{tab:Borncharges-AlN}. These charges are related to lattice vibrations and also directly related to the LO-TO splitting, as can be seen in Eqn.~(\ref{eq:D-corr}), where larger values for the Born effective charges provides a larger LO-TO splitting, especially in combination with smaller values of the dielectric tensor components. In the case of MgSiN$_{2}$ and MgGeN$_{2}$ the Born effective charge tensors are anisotropic with non-negligible off-diagonal elements. We note that the effective charges are sensitive to structural details. In a cubic environment, for example, the effective charges are diagonal and isotropic. However, for non-cubic structures and especially for systems with complex local environments, non-diagonal and anisotropic effective charges is not surprising.\cite{Cockayne2000} We note that the diagonal elements of the effective charge tensors are similar in MgSiN$_{2}$ and MgGeN$_{2}$, where the largest effective charges are found for the Si and Ge atoms in each system. We note that the average of the cation (and anion) diagonal components in MgSiN$_{2}$, e.g., ${\bm Z}^{\ast}_{xx}= ({\bm Z}^{\ast}_{xx}({\rm Mg})+{\bm Z}^{\ast}_{xx}({\rm Si}))/2=(1.92+3.17)/2=2.545$, are very similar to the effective charges in AlN shown in Table~\ref{tab:Borncharges-AlN}.

\section{$\Gamma$-point modes and Raman spectra}
\begin{figure}[t]
\includegraphics[width=8cm]{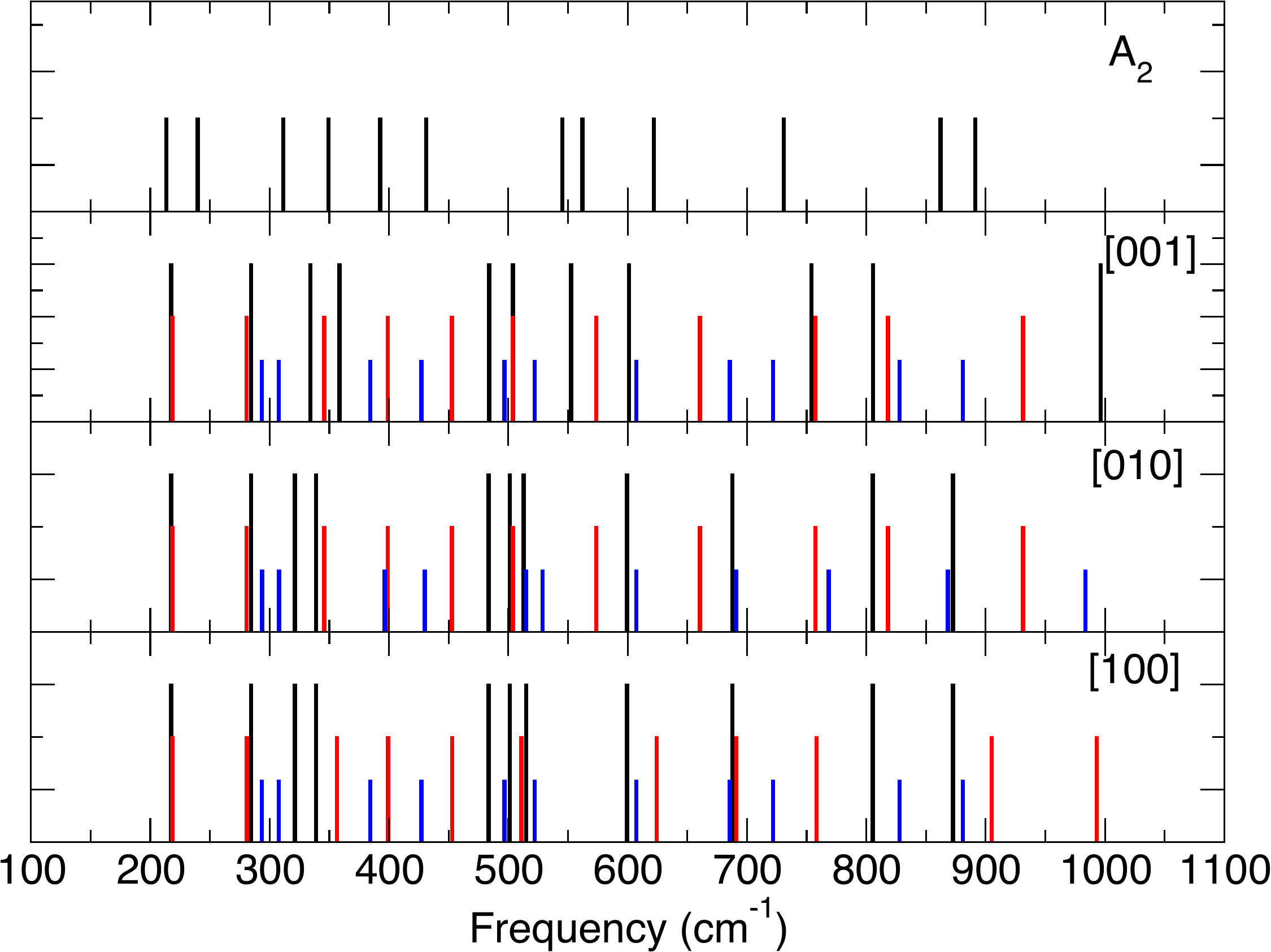}
\caption{\label{fig:gamma-mgsin2} (Color online) Calculated phonon modes at the $\Gamma$-point of MgSiN$_{2}$. The top panel contain the A$_{2}$ modes. The other panels contain the A$_{1}$, B$_{1}$ and B$_{2}$ modes for ${\bm q}\rightarrow0$ along [001], [010] and [100]. A$_{1}$, B$_{1}$ and B$_{2}$ modes are shown using (tall) black, red and (short) blue bars, respectively.}
\end{figure}
\begin{figure}[t]
\includegraphics[width=8cm]{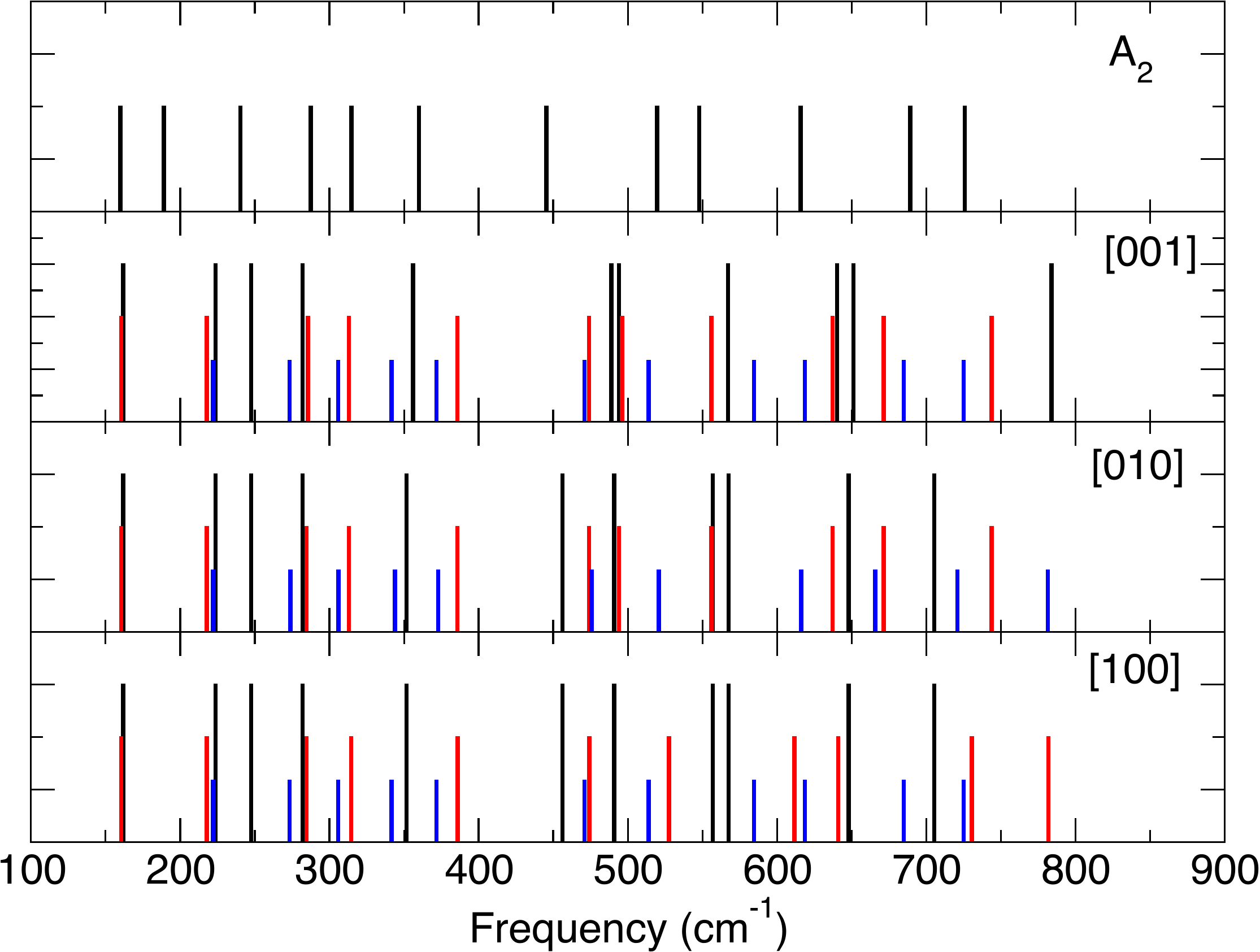}
\caption{\label{fig:gamma-mggen2} (Color online) Calculated phonon modes at the $\Gamma$-point of MgGeN$_{2}$. The top panel contain the A$_{2}$ modes. The other panels contain the A$_{1}$, B$_{1}$ and B$_{2}$ modes for ${\bm q}\rightarrow0$ along [001], [010] and [100]. A$_{1}$, B$_{1}$ and B$_{2}$ modes are shown using (tall) black, red and (short) blue bars, respectively.}
\end{figure}

We begin the discussion of the lattice dynamics by analysing the vibrations at the Brillouin zone center, i.e. at the $\Gamma$-point. The applicable point group for the orthorhombic Pna2$_{1}$ structure is $C_{2v}$ and it has four irreducible representations A$_{1}$, A$_{2}$, B$_{1}$ and B$_{2}$. We note that A$_{1}$, B$_{1}$ and B$_{2}$ correspond to modes with the same symmetry as a vector along the $z$, $x$ and $y$ crystal directions, respectively. These are the modes that are subject to a LO-TO splitting according to Eqn.~(\ref{eq:D-corr}), while the A$_{2}$ modes are unaffected. Since the orthorhombic unit cell contains 16 atoms there are 48 modes (12 for each symmetry), of which 3 are acoustic with zero frequency at the $\Gamma$-point. The zero frequency modes are of A$_{1}$, B$_{1}$ and B$_{2}$ symmetry. The remaining modes are all Raman active. Due to the LO-TO splitting the frequency of the remaining 11 A$_{1}$, 11 B$_{1}$ and 11 B$_{2}$ modes will depend on the direction from which the $\Gamma$-point is approached. That is, as an example, a static field in the $z$ direction will affect the A$_{1}$ modes, and therefore,
if the $\Gamma$-point is approached along the [001] direction, i.e. along the $z$-axis, the A$_{1}^{\rm LO}$ mode will be obtained. If the $\Gamma$-point is approached along $x$ or $y$ directions the A$_{1}^{\rm TO}$ mode will be obtained. In Figs.~\ref{fig:gamma-mgsin2} and \ref{fig:gamma-mggen2} we show the calculated frequencies of the $\Gamma$-point modes for both MgSiN$_{2}$ and MgGeN$_{2}$. For both systems the highest frequency mode along [001], [010] and [100] are A$_{1}^{\rm LO}$, B$_{2}^{\rm LO}$ and B$_{1}^{\rm LO}$ modes, respectively. In the case of MgSiN$_{2}$, the highest frequency is found to be 996.4~cm$^{-1}$ for the A$_{1}^{\rm LO}$, which is more than 100~cm$^{-1}$ higher than what we find for the E$_{1}^{\rm LO}$ mode (885.4~cm$^{-1}$) in wurtzite AlN. The A$_{1}^{\rm LO}$ mode in MgGeN$_{2}$ is 784.0~cm$^{-1}$. We note that in the Zn-based II-IV nitrides ZnSiN$_{2}$, ZnGeN$_{2}$ and ZnSnN$_{2}$ the highest frequencies are 980.0~cm$^{-1}$ (A$_{1}^{\rm LO}$),\cite{Paudel2007} 859.8~cm$^{-1}$ (B$_{2}^{\rm LO}$)\cite{Lambrecht2005} and 739~cm$^{-1}$ (B$_{2}^{\rm LO}$),\cite{Paudel2008} respectively. We therefore note that as the mass of the group II and IV elements increase the modes become softer and that both MgSiN$_{2}$ and ZnSiN$_{2}$ have modes with significantly higher frequency than found in AlN. 
\par
\begin{figure}[t]
\includegraphics[width=9cm]{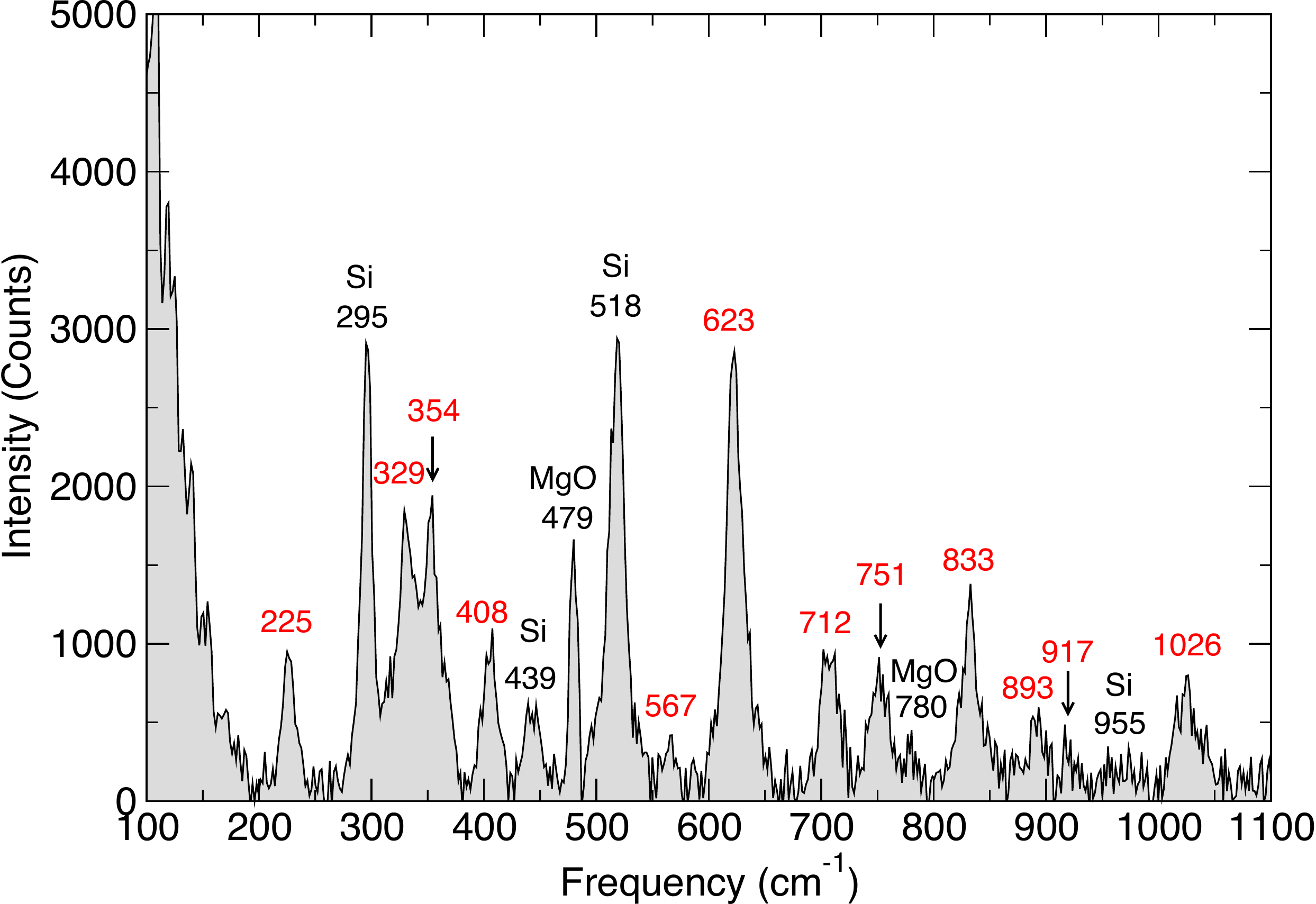}
\caption{\label{fig:raman} (Color online) Measured Raman spectra of a MgSiN$_{2}$ powder sample. Impurity peaks and assigned frequencies are shown in black. Unique peaks due to MgSiN$_{2}$ are shown in red.}
\end{figure}

Fig.~\ref{fig:raman} shows the results of a Raman measurement of a powder sample of MgSiN$_{2}$, where the principal MgSiN$_{2}$ peaks are highlighted with their frequency. We note that even though there exist 45 Raman active modes according to the symmetry analysis of MgSiN$_{2}$, all of them might not show up in a Raman measurement since the intensity of some of the peaks might be very low. In addition to the principal peaks, several peaks indicate the presence of free Si in the powder, which is in keeping with observations made for the Mg/Si$_{3}$N$_{4}$ synthesis route by Bruls {\it et al.}\cite{Bruls2002} Other impurity peaks corresponding to MgO are also present. The impurity peaks were assigned by referring to Raman spectra obtained for the impurity compounds on the same Raman system. As seen in Fig.~\ref{fig:raman}, there are 12 peaks derived solely from MgSiN$_{2}$ at 225, 329, 354, 408, 567, 623, 712, 751, 833, 893, 917 and 1026~cm$^{-1}$. These frequencies correspond agreeably with the calculated frequencies of Fig.~\ref{fig:gamma-mgsin2}. Furthermore, we find it to be likely that impurity peaks mask the presence of additional MgSiN$_{2}$ Raman peaks.

\section{Lattice dynamics}

\begin{figure}[t]
\includegraphics[width=8cm]{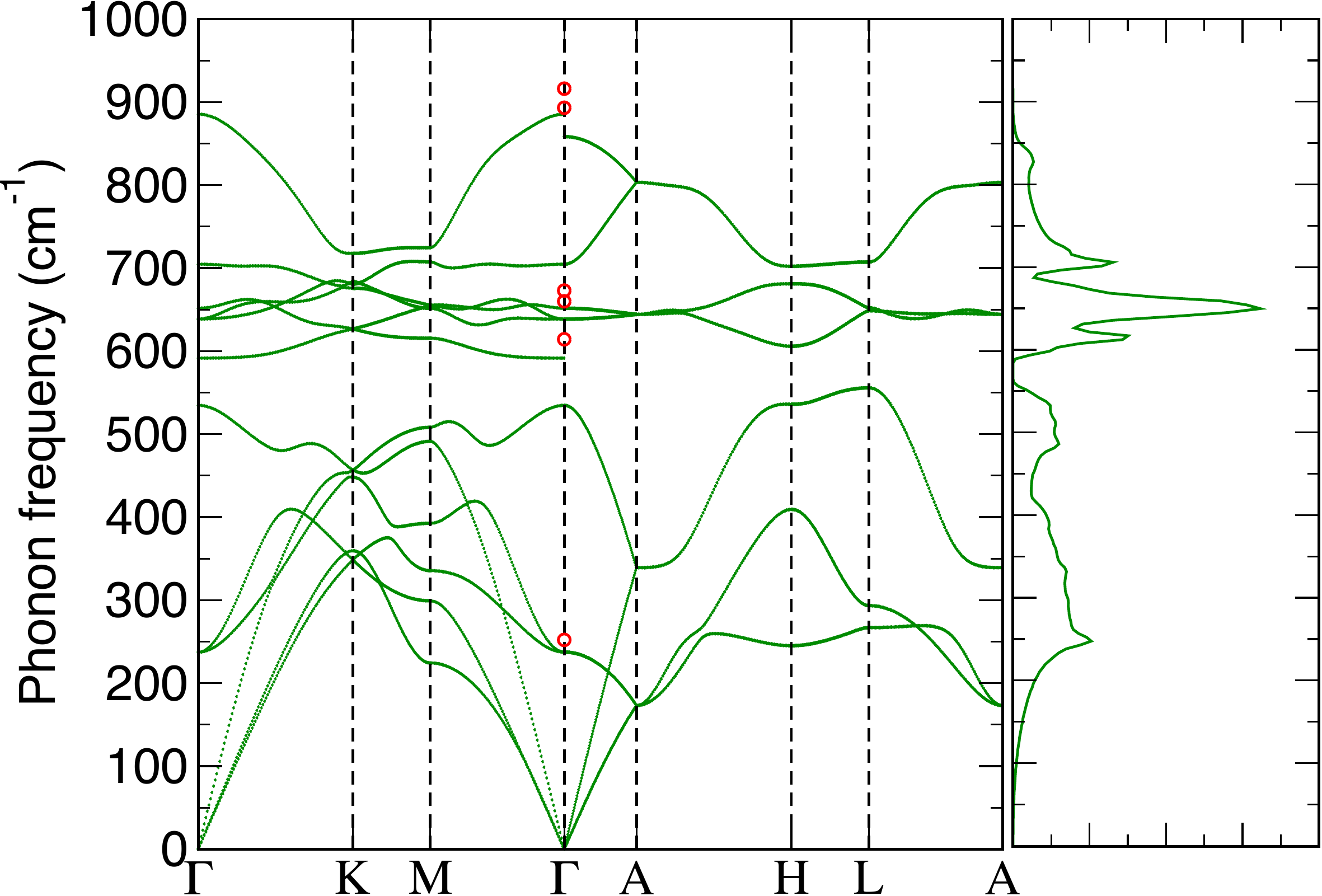}
\caption{\label{fig:phonons-AlN} (Color online) The calculated phonon dispersions along high symmetry directions in the Brillouin zone (left panel) and the phonon density of states (right panel) of wurtzite AlN. Experimental Raman frequencies\cite{McNeil1993} are shown using (red) circles.}
\end{figure}

Figure~\ref{fig:phonons-AlN} shows the phonon dispersions along high symmetry directions in the Brillouin zone of AlN in the wurtzite structure. The lattice dynamics of AlN has already been discussed thoroughly by, for example, Bungaro {\it et al.}\cite{Bungaro2000} and we will only summarise some general features of relevance when comparing to the more complex Mg-IV-N$_{2}$ systems.
The wurtzite unit cell contains 4 atoms and therefore there are 12 modes in the vibrational spectrum. There is a separation centered around 560 cm$^{-1}$ between higher optical modes and the lower energy modes. As was previously mentioned, the highest optical frequency is 885.4~cm$^{-1}$ for the LO mode with E$_{1}$ symmetry, which is slightly lower than the experimental frequency\cite{McNeil1993} of 916~cm$^{-1}$. We note that the underestimation obtained in the calculations compared to the available Raman frequencies in AlN, shown in Fig.~\ref{fig:phonons-AlN}, is likely due to the use of the PBE approximation, which tends to underestimate the phonon frequencies found experimentally due to the general underbinding obtained when using this approximation.
\par
\par
\begin{figure}[t]
\includegraphics[width=8cm]{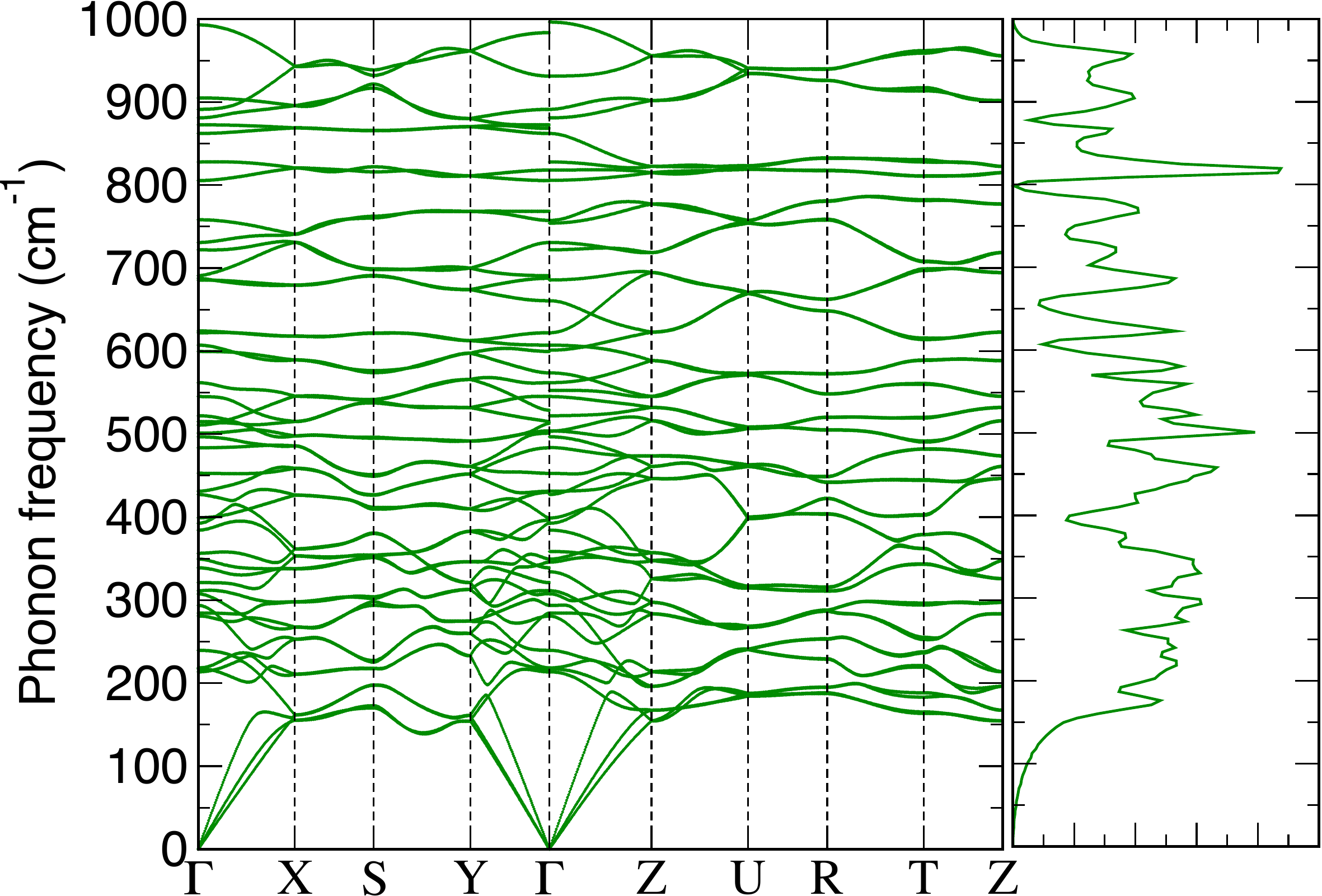}
\caption{\label{fig:phonons-MgSiN2} (Color online) The calculated phonon dispersions along high symmetry directions in the Brillouin zone (left panel) and the phonon density of states (right panel) of orthorhombic MgSiN$_{2}$.}
\end{figure}
\begin{figure}[t]
\includegraphics[width=8cm]{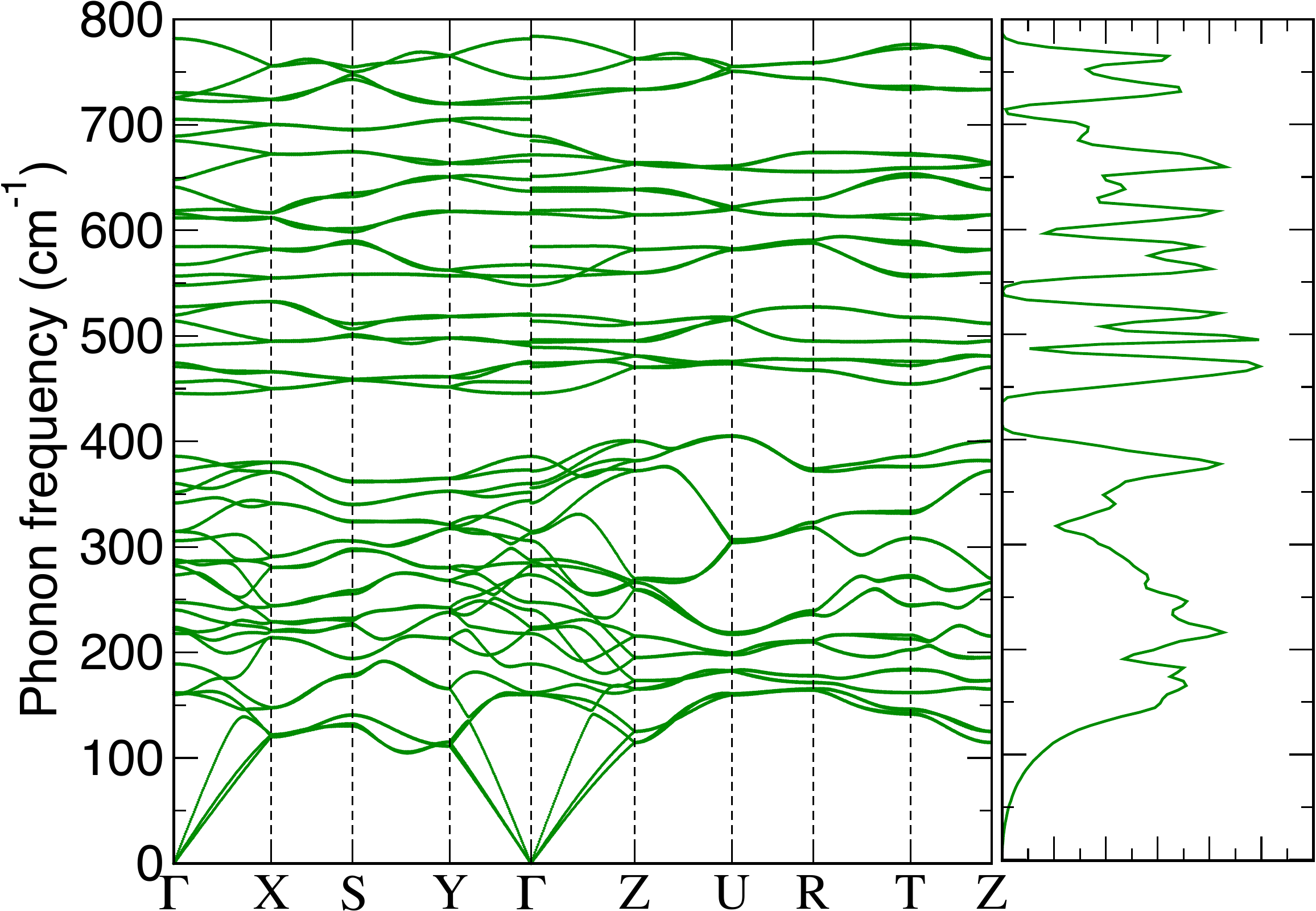}
\caption{\label{fig:phonons-MgGeN2} (Color online) The calculated phonon dispersions along high symmetry directions in the Brillouin zone (left panel) and the phonon density of states (right panel) of orthorhombic MgGeN$_{2}$.}
\end{figure}

By replacing two Al atoms for one Mg and one Si or Ge atom, the size of the system is doubled, but the unit cells for MgSiN$_{2}$ and MgGeN$_{2}$ contain 4 formula units (f.u.), as discussed in Section~\ref{sec:structure}, and the number of phonon modes is therefore 48. As a consequence, the phonon dispersions in MgSiN$_{2}$ and MgGeN$_{2}$ shown in Figs.~\ref{fig:phonons-MgSiN2} and \ref{fig:phonons-MgGeN2} are much more complex than the dispersions in AlN. In the case of MgSiN$_{2}$, we find that the separation found in AlN centered around 560 cm$^{-1}$ has vanished. Instead there is a small gap in the vibrations slightly below 800~cm$^{-1}$. In addition, there are a number of modes with small dispersion just above 800~cm$^{-1}$ that give rise to a tall peak in the phonon density of states shown in Fig.~\ref{fig:phonons-MgSiN2}. This feature is also found in ZnSiN$_{2}$,\cite{Paudel2008} even though the separation between the peak and the lower frequency modes is larger in the case of ZnSiN$_{2}$. There is, however, a significant difference between the vibrations in MgSiN$_{2}$ and ZnSiN$_{2}$: in the latter system there is a rather significant separation between the  modes below 200~cm$^{-1}$ and the higher frequency optical modes.\cite{Paudel2008} In MgSiN$_{2}$ these two regions have mixed into a single region. 
\par
When Si is substituted for the heavier Ge, the lattice vibrations become softer in general (see Fig.~\ref{fig:phonons-MgGeN2}), and for MgGeN$_{2}$ the highest frequency is 784.0~cm$^{-1}$. Compared to MgSiN$_{2}$, the lattice vibrations in MgGeN$_{2}$ are more structured with a clear separation between high frequency optical modes and lower frequency modes between 400~cm$^{-1}$ and 440~cm$^{-1}$. Furthermore, there are two additional gaps, where the lower gap is located at 530~cm$^{-1}$ and the other gap is located at about 710~cm$^{-1}$.
\begin{figure}[t]
\includegraphics[width=8cm]{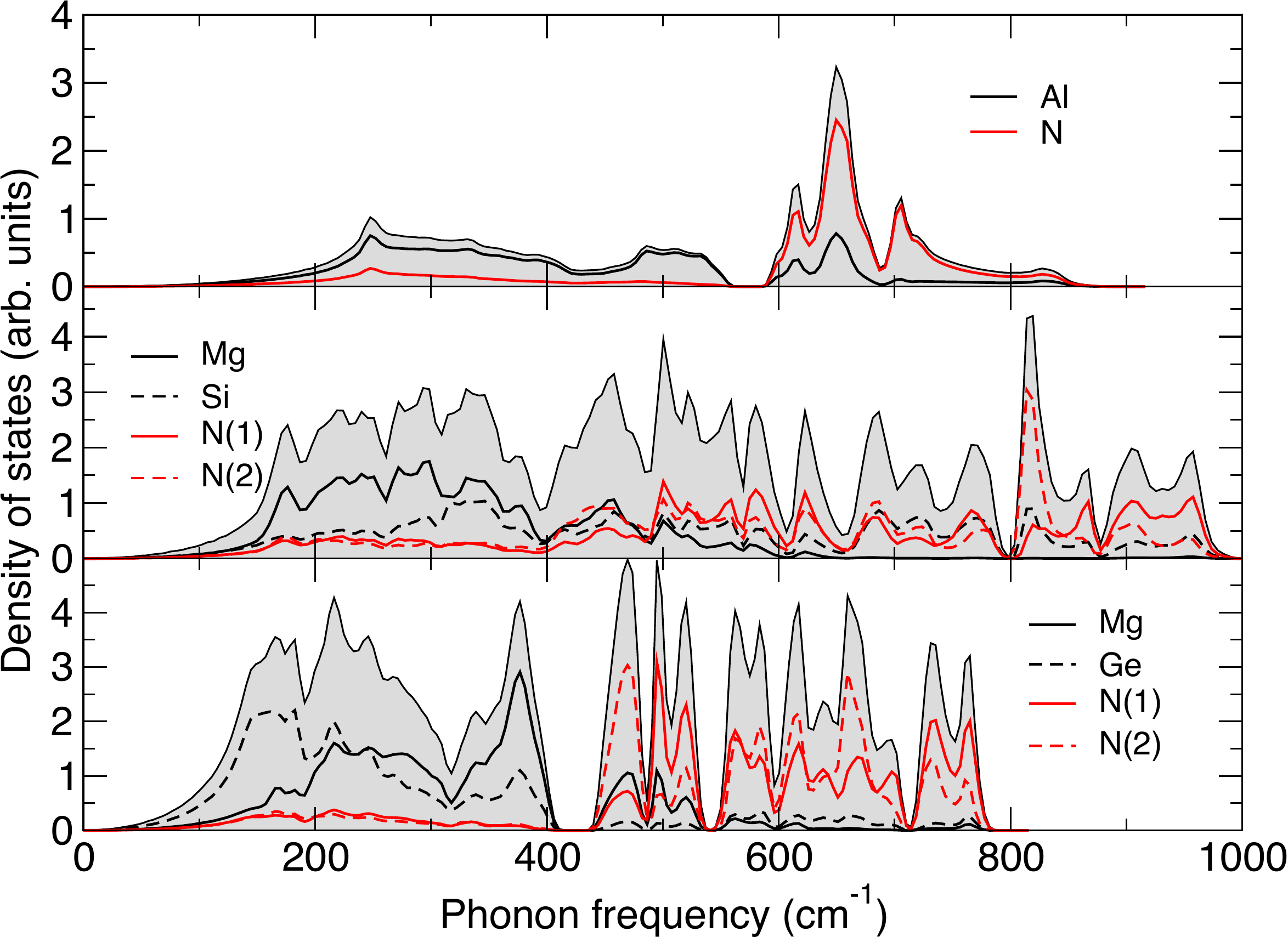}
\caption{\label{fig:phonon-dos} (Color online) The phonon density of states (DOS) and the projected DOS (pDOS) on to the atomic species of AlN, MgSiN$_{2}$ and MgGeN$_{2}$.}
\end{figure}
\par
In Fig.~\ref{fig:phonon-dos}, we show the phonon density of states of AlN, MgSiN$_{2}$ and MgGeN$_{2}$. In AlN it is clear that cation vibrations dominate at low frequencies and anion vibrations at high frequencies. The same is also true for MgSiN$_{2}$ and MgGeN$_{2}$, where in both cases the phonon density of states are dominated by cation vibrations up to about 400~cm$^{-1}$. Above this point, the vibrations are more dominated by nitrogen vibrations.

\section{Thermodynamic properties}
\begin{figure}[t]
\includegraphics[width=8cm]{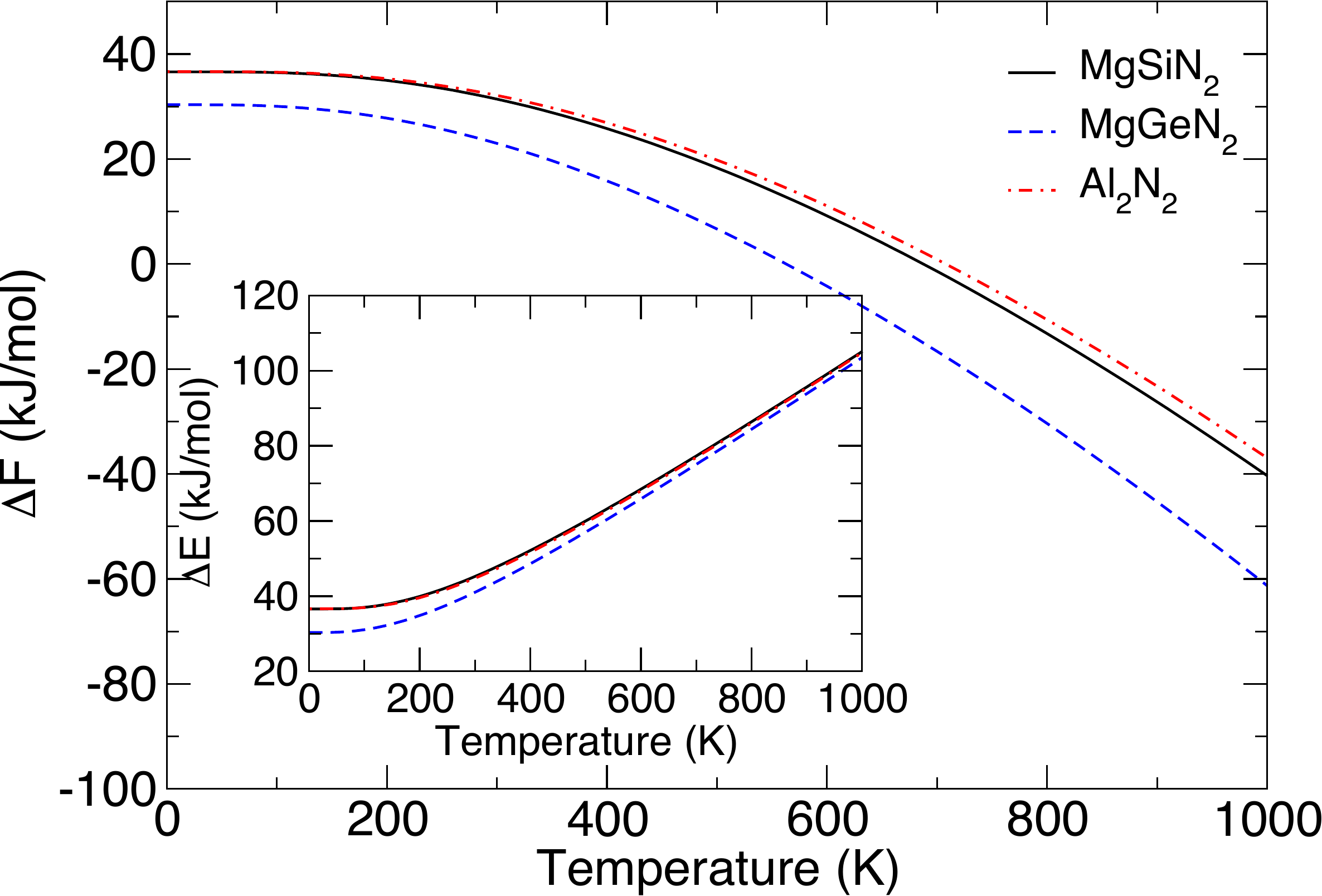}
\caption{\label{fig:energies} (Color online) The calculated Helmholtz free energy and total energy (inset) due to lattice vibrations of MgSiN$_{2}$, MgGeN$_{2}$ and AlN.}
\end{figure}

\begin{figure}[t]
\includegraphics[width=8cm]{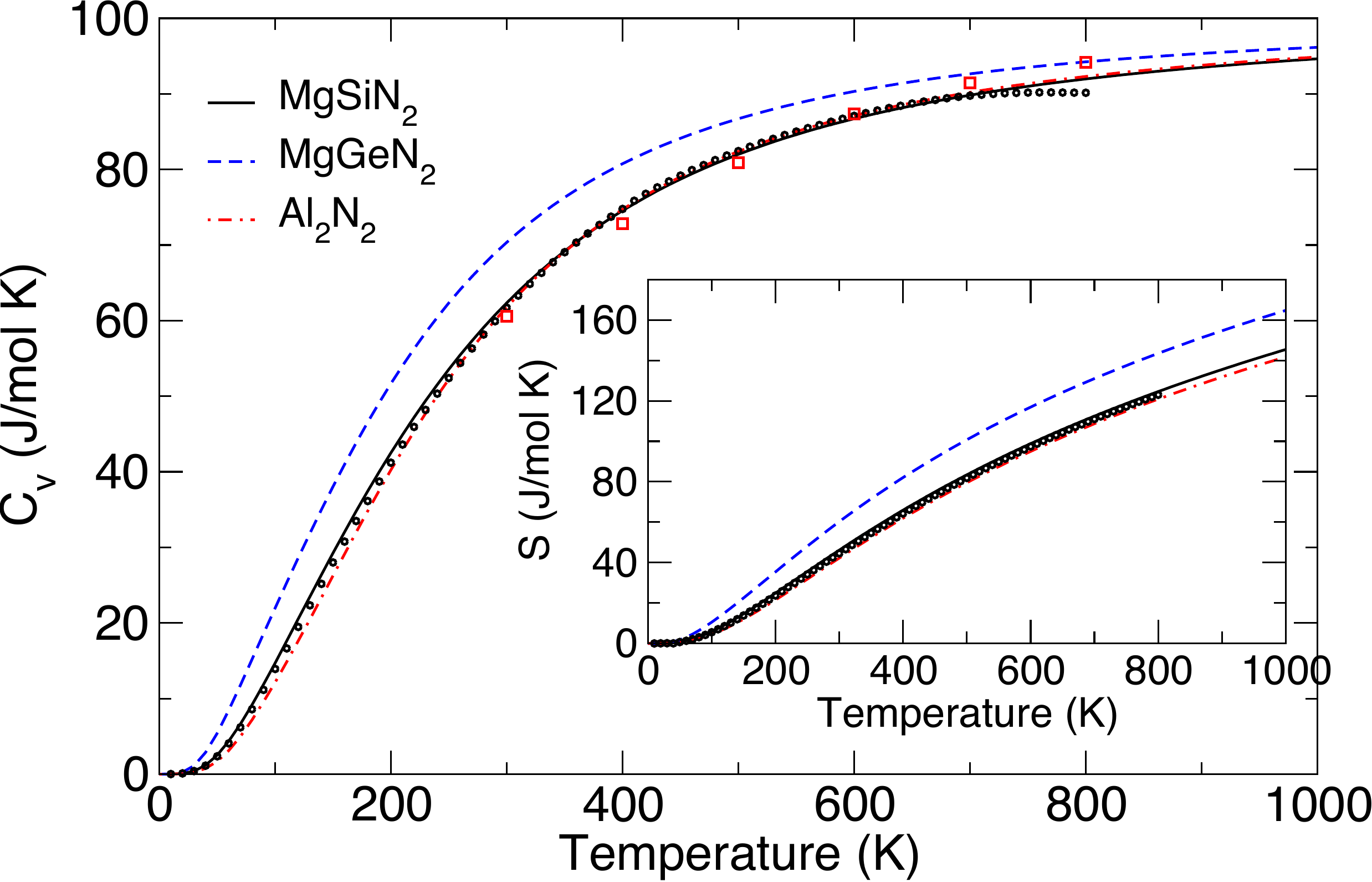}
\caption{\label{fig:specific-heat} (Color online) The specific heat at constant volume and entropy (inset) of MgSiN$_{2}$, MgGeN$_{2}$ and AlN. Experimental data for MgSiN$_{2}$ (black points) and AlN (red squares) are taken from Ref.~\onlinecite{Bruls1998}.}
\end{figure}

The contribution of the phonons to thermodynamic properties, such as the Helmholtz free energy $F$, the total energy $E$, the specific heat $C_{v} = (\partial E/\partial T)_{v}$ and the entropy $S = (E-F)/T$, is expressed through the partition function
\begin{equation}
Z = \Pi_{\lambda}\frac{e^{-\beta\hbar\omega_{\lambda}/2}}{1-e^{-\beta\hbar\omega_{\lambda}}},
\end{equation}
where $\lambda$ is a combined summation index over both modes and q-points and $\omega_{\lambda}$ is the vibrational frequency of mode $\nu$ at q-point $\bm q$. The Helmholtz free energy is given by
\begin{equation}
F = -\frac{1}{\beta}{\rm ln}\,Z
= \sum_{\lambda} \left[\frac{\hbar\omega_{\lambda}}{2} + \frac{1}{\beta}{\rm ln}\,\bigl( 1-{\rm exp}\bigl(-\beta\hbar\omega_{\lambda}\bigr)\bigr)\right]
\end{equation}
and the total energy is given by
\begin{equation}
E = -\frac{\partial{\rm ln}\,Z}{\partial\beta}
 = \sum_{\lambda} \hbar\omega_{\lambda}\left[ \frac{1}{2} + \frac{1}{{\rm exp}(\beta\hbar\omega_{\lambda})-1}\right].
\end{equation}
\par
In Figs.~\ref{fig:energies} and \ref{fig:specific-heat} we show the calculated free energies, total energies, heat capacity and entropy of MgSiN$_{2}$, MgGeN$_{2}$ and AlN. Note that in order to facilitate an easier comparison between AlN and the two II-IV nitrides, we have plotted the AlN related curves as if the system is twice as large, i.e. we show the free energy, total energy, heat capacity and entropy for Al$_{2}$N$_{2}$ instead of AlN. Otherwise, all properties shown in Figs.~\ref{fig:energies} and \ref{fig:specific-heat} are about twice as large for MgSiN$_{2}$ than for AlN.
\par
The zero point energies, i.e., the total energy at $T=0$~K, for MgSiN$_{2}$, MgGeN$_{2}$ and AlN are 36.6, 30.3 and 18.3 kJ/mol, respectively, or 0.190, 0.157 and 0.190 eV per nitrogen atom, respectively. In comparison, the calculated zero point energy in ZnSiN$_{2}$ is 34.0 kJ/mol or 0.176 eV per nitrogen atom,\cite{Paudel2008} which is slightly less than what is found for MgSiN$_{2}$ but slightly higher than what is found for MgGeN$_{2}$. This is expected due to the differences in the mass between Mg, Si, Ge and Zn, where Zn is heavier than Mg but Ge is heavier than Si and Zn which makes the MgGeN$_{2}$ a heavier system than both MgSiN$_{2}$ and ZnSiN$_{2}$. 
\par
The curves for AlN and MgSiN$_{2}$ follow each other closely in Figs.~\ref{fig:energies} and \ref{fig:specific-heat}. The free and total energies of MgGeN$_{2}$ are smaller than those of MgSiN$_{2}$; the difference between the free energy of the two compounds increases with temperature, but in contrast the total energy reduces with increasing temperature. In Fig.~\ref{fig:specific-heat}, it is clear that both the specific heat and entropy are larger in MgGeN$_{2}$ compared to MgSiN$_{2}$. This behaviour is identical to what is observed for Zn-IV-N$_{2}$, where the free and total energies decrease from ZnSiN$_{2}$ to ZnGeN$_{2}$ to ZnSnN$_{2}$ while the heat capacity and entropy increase from ZnSiN$_{2}$ to ZnGeN$_{2}$ to ZnSnN$_{2}$.\cite{Paudel2008}
\par
When comparing our calculated thermodynamic properties to available experimental studies, we find a very good agreement regarding the vibrational entropy in the case of MgSiN$_{2}$, as is shown in Fig.~\ref{fig:specific-heat}. In Fig.~\ref{fig:specific-heat} we also show the experimental heat capacity at constant pressure ($C_{p}$), which is related to the heat capacity at constant volume ($C_{v}$) by
\begin{equation}\label{eq:cp-cv}
C_{p}-C_{v} = 9\alpha^2\frac{V_{m}T}{\beta_{T}},
\end{equation}
where $\alpha$ is the thermal expansion, $V_{m}$ is the molar volume and $\beta_{T}$ is the isothermal compressibility. The difference between $C_{p}$ and $C_{v}$ is a continuously increasing function of the temperature. By using experimental values for the thermal expansion, volume and compressibility, Bruls {\it et al.}\cite{Bruls1998} have shown that the difference in the left hand side of Eqn.~(\ref{eq:cp-cv}) at 800~K is 1.2 J$\cdot$mol$^{-1}\cdot$K$^{-1}$, i.e. a relative difference of about 1.3~\%, which was well inside the accuracy of the measurement. Therefore, for sufficiently low temperatures the heat capacities at constant pressure and constant volume are approximately the same for MgSiN$_{2}$ and a comparison between experimental values for the heat capacity at constant pressure and the theoretical heat capacity at constant volume is relevant. As can be seen in Fig.~\ref{fig:specific-heat} the experimental and theoretical curves follow each other closely up to about 700~K.

\section{Summary and conclusions}

We have performed density functional calculations of the structural and lattice dynamical properties of MgSiN$_{2}$ and MgGeN$_{2}$. Our calculations are in very good agreement with available experimental results, especially regarding the structural properties, but also regarding thermodynamic properties, i.e., heat capacity and entropy, in the case of MgSiN$_{2}$. We find that MgSiN$_{2}$ is structurally very similar to wurtzite AlN, with a very good in-plane lattice matching, while MgGeN$_{2}$ has a larger volume and more similar to wurtzite GaN. The phonon dispersions in MgSiN$_{2}$ and MgGeN$_{2}$ are much more complex than in AlN. We also find that the highest optical mode is about 100~cm$^{-1}$ higher in energy in MgSiN$_{2}$ compared to AlN and that the lattice vibrations in MgGeN$_{2}$ are softer than in both MgSiN$_{2}$ and AlN. The free energy and total energy due to lattice vibrations are smaller in MgGeN$_{2}$ than in MgSiN$_{2}$, while the opposite is the case for the heat capacity and the entropy. When comparing the thermodynamic properties, i.e. free energy, total energy, heat capacity and entropy, of MgSiN$_{2}$ and AlN, we find that the curves as a function of temperature follow each other closely with the difference being an approximate factor of 2 per mole. For example, the zero point energy in MgSiN$_{2}$ is 36.6 kJ/mol while it is 18.3 kJ/mol in AlN.

\section{Acknowledgements}
We acknowledge support from the Leverhulme Trust via M. A. Moram's Research Leadership Award (RL-007-2012). M. A. Moram further acknowledges support from the Royal Society through a University Research Fellowship. This work used the Imperial College high performance computing facilities and, via our membership of the UK's HEC Materials Chemistry Consortium funded by EPSRC (EP/L000202), the ARCHER UK National Supercomputing Service (http://www.archer.ac.uk).

\bibliography{mgsin2}

\end{document}